\documentclass[review]{elsarticle}

\usepackage{amssymb}
\usepackage{amsmath}
\usepackage{amsfonts}
\usepackage{amsbsy}
\usepackage{graphicx}
\usepackage{float}
\usepackage{color}

\newcommand{\pp}{{\mbox{\boldmath $p$}}}
\newcommand{\qq}{{\mbox{\boldmath $q$}}}
\newcommand{\kk}{{\mbox{\boldmath $k$}}}
\newcommand{\xx}{{\mbox{\boldmath $x$}}}
\newcommand{\xxb}{{\mbox{\boldmath $\overline{x}$}}}
\newcommand{\pint}{\int^{\!-}_{\!\!-}\!\!\!\!\!\int}
\newcommand{\bq}{\begin{eqnarray}}
\newcommand{\eq}{\end{eqnarray}}
\newcommand{\tpi}{\tilde{\pi}}
\newcommand{\cO}{{\cal O}}

\newcommand{\bc}{\begin{center}}
\newcommand{\ec}{\end{center}}
\newcommand{\ba}{\begin{array}{ll}}
\newcommand{\ea}{\end{array}}

\newcounter{ctr}
\newcounter{sbqi}
\def\lista{\begin{list}{(\roman{ctr})}{\usecounter{ctr}
\setlength{\rightmargin}{0.0truecm}\setlength{\leftmargin}{\labelwidth}}}

\begin{document}
\title{Low temperature acoustic polaron localization}  

\author[els]{Riccardo Fantoni}
\ead{rfantoni@ts.infn.it}
\address[els]{Dipartimento di Scienze dei Materiali e Nanosistemi,
  Universit\`a Ca' Foscari Venezia, Calle Larga S. Marta DD2137, I-30123
  Venezia, Italy}

\date{\today}

\begin{abstract}
We calculate the properties of an acoustic polaron in three dimensions
in thermal equilibrium at a given low temperature using 
the path integral Monte Carlo method. The specialized numerical method
used is described in full details, thus complementing our previous paper
[R. Fantoni, Phys. Rev. B {\bf 86}, 144304 (2012)], and it appears to
be the first time it has been used in this context. Our results are in
favor of the presence of a phase transition from a localized state to
an extended state for the electron as the phonon-electron coupling
constant decreases. The phase transition manifests itself with a jump
discontinuity in the potential energy as a function of the coupling
constant and it affects the properties of the path of the electron in
imaginary time: In the weak coupling regime the electron is in an
extended state whereas in the strong coupling regime it is found in a 
self-trapped state.
\end{abstract}


\maketitle
\section{Introduction}
\label{sec:introduction}

An electron in a ionic crystal polarizes the lattice in its
neighborhood. An electron moving with its accompanying
distortion of the lattice has sometimes been called a ``polaron''
\cite{Gerlach1991,Devresee2009}. 
Since 1933 Landau addresses the possibility whether an electron can be
self-trapped (ST) in a deformable lattice
\cite{Landau1933,Landau1946,Landau1948}. This fundamental problem in
solid state physics has been intensively studied for an optical
polaron in an ionic crystal
\cite{Frohlich1950,Frohlich1954,Feynman1955,Peeters1982,Mason1986,Mitra1987}. 
Bogoliubov approached the polaron strong coupling limit with one of
his canonical transformations. Feynman used his path integral
formalism and a variational principle to develop an all coupling
approximation for the polaron ground state \cite{Feynman}. Its
extension to finite temperatures appeared first 
by Osaka \cite{Osaka1959,Osaka1965}, and more recently by Castrigiano
{\it et al.} \cite{Castrigliano1983,Castrigliano1984,Khandekar1986}. 
Recently the polaron problem has gained new interest as it
could play a role in explaining the properties of the high
$T_c$ superconductors \cite{Sheng}. The polaron problem has also been
studied to describe an impurity 
in a Bose-Einstein ultracold quantum gas condensate of atoms
\cite{Tempere2009}. In this context evidence for a transition
between free and self-trapped optical polarons is found. For the
solid state optical polaron no ST state has been found 
yet \cite{Feynman1955,Peeters1982,Mitra1987}. 

The acoustic modes of lattice vibration are known to be responsible
for the appearance of the ST state
\cite{Toyozawa1961,Kuper1963,Gerlach1991}. Contrary to 
the optical mode which interacts with the electron through Coulombic
force and is dispersionless, the acoustic phonons have a linear
dispersion coupled to the electron through a short range potential
which is believed to play a crucial role in forming the ST
state \cite{Peteers1985}. Acoustic modes have also been widely
studied \cite{Gerlach1991}. Sumi and Toyozawa 
generalized the optical polaron model by including a coupling to the
acoustic modes \cite{sumi73}. Using Feynman's variational approach,
they found that the electron is ST with a very large effective mass as
the acoustic coupling exceeds a critical value. Emin and Holstein also
reached a similar conclusion within a scaling theory \cite{Emin1976} in
which the Gaussian trial wave function is essentially identical to the
harmonic trial action used in the Feynman's variational approach in
the adiabatic limit \cite{Fisher1986}. 

The ST state distinguishes itself from an extended state (ES) where
the polaron has lower mass and a bigger radius. A polaronic phase
transition separates the two states with a breaking of translational
symmetry in the ST one \cite{Gerlach1991}. The variational approach is
unable to clearly assess the existence of the phase transition
\cite{Gerlach1991}. In particular Gerlach and L\"owen
\cite{Gerlach1991} concluded that no phase transition exists in a
large class of polarons. The three dimensional acoustic polaron is not
included in the class but Fisher {\it et al.} \cite{Fisher1986} argued
that its ground state is delocalized.  

In a recent work \cite{Fantoni2012} we employed for the first
time a specialized path integral 
Monte Carlo (PIMC) method \cite{ceperley95,Pierleoni2001} to the
continuous, highly non-local,  
acoustic polaron problem at low temperature which is valid at all
values of the coupling 
strength and solves the problem exactly (in a Monte Carlo sense). The
method differs from previously employed methods
\cite{Alexandrou1990,Alexandrou1992b,Crutz1981,Takahashi1983,Wang1998,Kornilovitch1997,Kornilovitch2007}
and hinges on the L\'evy construction and the multilevel Metropolis
method with correlated sampling. In such work the potential energy was
calculated and it was 
shown that like the effective mass it usefully signals the transition
between the ES and the ST state. Properties of ES and ST states were
explicitly shown through the numerical simulation.

Aim of the present paper is to give a detailed description of
the PIMC method used in that calculation and some additional
numerical results in order to complement the brief paper of Ref.
\cite{Fantoni2012}. 

The work is organized as follows: in section \ref{sec:model} we
describe the acoustic polaron model and Hamiltonian, in section
\ref{sec:observables} we describe the observables we are going to
compute in the simulation, in section \ref{sec:pi} we describe the
PIMC numerical scheme employed, in section \ref{sec:path} we
describe the multilevel Metropolis method for sampling the path, in
section \ref{sec:metropolis} we describe the choice of the transition
probability and the level action, in section \ref{sec:sampling} we
describe the correlated sampling. Section \ref{sec:results} is for the
results, and section \ref{sec:conclusions} is for final remarks.

\section{The model}
\label{sec:model}
The acoustic polaron can be described by the following quasi-continuous
model \cite{Frohlich1954,sumi73},
\bq
\hat{H}=\frac{\hat{\pp}^2}{2m}+\sum_{\kk}\hbar\omega_k
\hat{a}_{\kk}^{\dagger}\hat{a}_{\kk} + \sum_{\kk} \left(i \Gamma_k
\hat{a}_{\kk} e^{i\kk \hat{\xx}} + \mbox{H.c.}\right)~~.
\eq
Here $\hat{\xx}$ and $\hat{\pp}$ are the electron coordinate and momentum
operators respectively and $\hat{a}_\kk$ is the annihilation operator 
of the acoustic phonon with wave vector $\kk$. The first term in the
Hamiltonian is the kinetic energy of the electron, the second term the
energy of the phonons and the third term the coupling energy between
the electron and the phonons.
The electron coordinate $\xx$ is a continuous variable, while 
the phonons wave vector $\kk$ is restricted by the Debye cut-off $k_o$. 
The acoustic phonons have a dispersion relation $\omega_k=uk$ ($u$ 
being the sound velocity) and they interact with the electron of mass $m$
through the interaction vertex $\Gamma_k=\hbar u k_o (S/N)^{1/2}
(k/k_o)^{1/2}$ according to the deformation potential analysis of
Ref. \cite{Bardeen1950}.
$S$ is the coupling constant between the electron and the phonons and 
$N$ the number of unit cells in the crystal with $N/V=(4\pi/3)
(k_o/2\pi)^3$ by Debye approximation and $V$ the crystal volume.

Using the path integral representation (see Ref. \cite{Feynman}
section 8.3), the phonon part in the
Hamiltonian can be exactly integrated owing to its quadratic form in
phonon coordinates, and one can write the partition
function for a polaron in thermal equilibrium at an absolute
temperature $T$ ($\beta=1/k_BT$, with $k_B$ Boltzmann constant) as
follows, 
\bq
Z=\int d\xx \pint_{\xx=\xx(0)}^{\xx=\xx(\hbar\beta)} e^{
-\frac{1}{\hbar}{\cal S}[\xx(t),\dot{\xx}(t),t]} {\cal D}\xx(t)~~,
\eq 
where the action ${\cal S}$ is given by \cite{feynman65},\footnote{
This is an approximation as $e^{-\beta\omega_k}$ is neglected. The
complete form is obtained by replacing $e^{-\omega_k|t-s|}$ by
$e^{-\omega_k|t-s|}/(1-e^{-\beta\omega_k})+
e^{\omega_k|t-s|}e^{-\beta\omega_k}/(1-e^{-\beta\omega_k})$.
But remember that $\beta$ is large.}
\bq \nonumber
{\cal S}&=&\frac{m}{2}\int_0^{\hbar\beta}\dot{\xx}^2(t)dt-\frac{1}{2\hbar}
\int_0^{\hbar\beta}dt\int_0^{\hbar\beta}ds \int\frac{d\kk}{(2\pi)^3} 
\Gamma_k^2 e^{i\kk\cdot(\xx(t)-\xx(s))-\omega_k|t-s|}\\
&=&{\cal S}_f+{\cal U}~~.
\eq
Here ${\cal S}_f$ is the {\it free particle action}, and ${\cal U}$ the 
{\it inter-action} and we denoted with a dot a time derivative as
usual. Using dimensionless units $\hbar=m=uk_o=k_B=V=1$ the action becomes, 
\bq \label{action}
{\cal S}=\int_0^{\beta}\frac{\dot{\xx}^2(t)}{2}dt+
\int_0^{\beta}dt\int_0^{\beta}ds \,V_{eff}(|\xx(t)-\xx(s)|,|t-s|)~~,
\eq 
with the electron moving subject to an effective retarded potential,
\bq
V_{eff}&=&-\frac{S}{2I_D}\int_{q\leq 1}d\qq\, q e^{i\sqrt{\frac{2}{\gamma}}
\qq\cdot(\xx(t)-\xx(s))-q|t-s|}\\ \label{veff}
&=&-\frac{3S}{2}\sqrt{\frac{\gamma}{2}}\frac{1}{|\xx(t)-\xx(s)|}\int_0^1
dq\,q^2\sin\left(\sqrt{\frac{2}{\gamma}}q|\xx(t)-\xx(s)|\right)e^{-q|t-s|}~~,
\eq
where $\qq=\kk/k_o$, $I_D=\int_{q\leq 1} d\qq=4\pi/3$, and we have 
introduced a non-adiabatic parameter $\gamma$ defined as the ratio of the 
average phonon energy, $\hbar u k_o$ to the electron band-width, 
$(\hbar k_o)^2/2m$. This parameter is of order of $10^{-2}$ in typical
ionic crystals with broad band so that the ST state is well-defined
\cite{sumi73}. In our simulation we took $\gamma=0.02$. Note that the
integral in (\ref{veff}) can be solved analytically and the resulting
function tabulated.

\section{The observables}
\label{sec:observables}
In particular the internal energy $E$ of the polaron is given by, 
\bq
E=-\frac{1}{Z}\frac{\partial Z}{\partial \beta}=\frac{1}{Z}\int d\xx
\pint e^{-{\cal S}}\frac{\partial {\cal S}}{\partial \beta}{\cal D}\xx=\left
\langle\frac{\partial {\cal S}}{\partial \beta}\right\rangle~~,
\eq 
where the internal energy tends to the ground state energy
in the large $\beta\longrightarrow \infty$ limit. 

Scaling the Euclidean time $t=\beta t^\prime$ and $s=\beta s^\prime$
in Eq. (\ref{action}), deriving ${\cal S}$ with respect to
$\beta$, and undoing the scaling, we get,
\bq \nonumber
\frac{\partial {\cal S}}{\partial \beta}=-\frac{1}{\beta}\int_0^\beta\frac{
\dot{\xx}^2}{2}dt-\frac{S}{2I_D}\int_0^\beta dt\int_0^\beta ds \times\\
\int_{q\leq 1}d\qq\, q e^{i\sqrt{\frac{2}{\gamma}}
\qq\cdot(\xx(t)-\xx(s))-q|t-s|}
\frac{1}{\beta}(2-q|t-s|)~~,
\eq
where the first term is the kinetic energy contribution to the internal energy,
${\cal K}$, and the last term is the potential energy contribution,
${\cal P}$,
\bq \nonumber
{\cal P}=-\frac{3S}{2\beta}\int_0^\beta dt\int_0^\beta ds\int_0^1 dq\,q^3
\frac{\sin\left(\sqrt{\frac{2}{\gamma}}q|\xx(t)-\xx(s)|\right)}
{\sqrt{\frac{2}{\gamma}}q|\xx(t)-\xx(s)|}e^{-q|t-s|}\times\\
(2-q|t-s|)~~.
\eq
So that,
\bq
E=\langle {\cal K}+{\cal P} \rangle~~.
\eq 

An expression for ${\cal K}$ not involving the polaron speed, can be obtained 
by taking the derivative with respect to $\beta$ after having scaled both 
the time, as before, and the coordinate $\xx=\sqrt{\beta}\xx^\prime$. 
Undoing the scaling in the end one gets,
\bq \nonumber
{\cal K}&=&-\frac{S}{4\beta I_D}\int_0^\beta dt\int_0^\beta ds
\int_{q\leq 1}d\qq\, q e^{i\sqrt{\frac{2}{\gamma}}\qq\cdot(\xx(t)-\xx(s))-q|t-s|}
\times \\
&&\left[i\sqrt{\frac{2}{\gamma}}
\qq\cdot(\xx(t)-\xx(s))\right]\\ \nonumber
&=&-\frac{3S}{4\beta}\int_0^\beta dt\int_0^\beta ds\int_0^1 dq\,q^3\left[
\cos\left(\sqrt{\frac{2}{\gamma}}q|\xx(t)-\xx(s)|\right)-\right.\\
&&\left.
\frac{\sin\left(\sqrt{\frac{2}{\gamma}}q|\xx(t)-\xx(s)|\right)}
{\sqrt{\frac{2}{\gamma}}q|\xx(t)-\xx(s)|}\right]e^{-q|t-s|}~~.
\eq
In the following we will explain how we calculated the potential
energy $P=\langle{\cal P}\rangle$.
\section{Discrete path integral expressions}
\label{sec:pi}
Generally we are interested in calculating the density matrix $\hat{ 
\rho}=\exp(-\beta\hat{H})$ in the electron coordinate basis, namely,
\bq 
\rho(\xx_a,\xx_b;\beta)=\pint_{\xx=\xx_a}^{\xx=\xx_b}e^{-{\cal S}}{\cal D}
\xx(t)~~.
\eq

To calculate the path integral, we first choose a 
subset of all paths. To do this ,we divide the independent variable,
Euclidean time, into {\it steps} of width
\bq
\tau = \beta/M~~.
\eq
This gives us a set of {\it times}, $t_k=k\tau$ spaced a distance $\tau$
apart between $0$ and $\beta$ with $k=0,1,2,\ldots,M$.

At each time $t_k$ we select the special point $\xx_k=\xx(t_k)$, the $k^{th}$
{\it time slice}. We construct a path by connecting all points so selected 
by straight lines. It is possible to define a sum over all paths constructed
in this manner by taking a multiple integral over all values of $\xx_k$
for $k=1,2,\ldots,M-1$ where $\xx_0=\xx_a$ and $\xx_M=\xx_b$ are the two 
fixed ends. The resulting equation is, 
\bq \label{path-integral}
\rho(\xx_a,\xx_b;\beta)=\lim_{\tau\rightarrow 0}\frac{1}{A}
\int_{-\infty}^\infty\int_{-\infty}^\infty\cdots\int_{-\infty}^\infty
e^{-{\cal S}}\frac{d\xx_1}{A}\cdots \frac{d\xx_{M-1}}{A}~~,
\eq   
where the normalizing factor $A=(2\pi\tau)^{3/2}$.

The simplest discretized expression for the action can then be written 
as follows,
\bq \label{s-discr}
{\cal S}=\sum_{k=1}^M \frac{(\xx_{k-1}-\xx_k)^2}{2\tau}+\tau^2\sum_{i=1}^M
\sum_{j=1}^M V(t_i,t_j)~~,
\eq 
where $V(t_i,t_j)=V_{eff}(|\xx_i-\xx_j|,|i-j|)$ is a symmetric
two variables function, $V(s,t)=V(t,s)$. In our simulation we tabulated 
this function taking $|\xx_i-\xx_j|=0,0.1,0.2,\ldots,10$ and $|i-j|=
0,1,\ldots,M$.

In writing Eq. (\ref{s-discr}) we used the following approximate expressions,
\bq \label{d-1}
&&\dot{\xx}_k=\frac{\xx_k-\xx_{k-1}}{\tau}+O(\tau)~~,\\ \label{d-2} 
&&\int_{t_{k-1}}^{t_k}\dot{\xx}^2(t)\,dt=\dot{\xx}_k^2\tau+O(\tau^2)~~,\\ 
\label{d-3} 
&&\int_{t_{i-1}}^{t_i}\int_{t_{j-1}}^{t_j}V(s,t)\,dsdt=V(t_i,t_j)\tau^2
+O(\tau^3)~~.
\eq
If we take $V=0$ in Eq. (\ref{s-discr}) the $M-1$ Gaussian integrals in
(\ref{path-integral}) can be done analytically.  
The result is the exact free particle density matrix,
\bq
\rho_f(\xx_a,\xx_b;\beta)=(2\pi\beta)^{-3/2}e^{\frac{1}{2\beta}(\xx_a-
\xx_b)^2}~~.
\eq
Thus approximations (\ref{d-1}) and (\ref{d-2}) allow us to rewrite the 
polaron density matrix as follows,
\bq \nonumber
\rho(\xx_a,\xx_b;\beta)=\int\cdots\int d\xx_1\cdots d\xx_{M-1}
\,\rho_f(\xx_a,\xx_1;\tau)\cdots \rho_f(\xx_{M-1},\xx_M;\tau)\times\\
e^{\tau^2\sum_i\sum_jV(t_i,t_j)}~~.
\eq
In the next section we will see that this expression offers a useful 
starting point for the construction of an algorithm for the sampling
of the path: the L\'{e}vy construction and the analogy with classical
polymer systems or the classical isomorphism described in
\cite{ceperley95}).  

The partition function is the trace of the density matrix,
\bq
Z=\int d\xx\, \rho(\xx,\xx;\beta)~~.
\eq
This restrict the path integral to an integral over closed paths only.
In other words the paths we need to consider in calculating $Z$ (and 
hence $F$) are closed by the {\it periodic boundary condition},
$\xx_M=\xx_0=\xx$.

To calculate the internal energy we need then to perform the following $M$
dimensional integral,
\bq
E=\left.\frac{1}{Z}\int_{-\infty}^\infty\int_{-\infty}^\infty\cdots
\int_{-\infty}^\infty d\xx_0d\xx_1\cdots d\xx_{M-1}\,e^{-{\cal S}}({\cal
  P}+{\cal K})\right|
_{\xx_M=\xx_0}~~.
\eq 
To do this integral we use the Monte Carlo simulation technique
described next.

\section{Sampling the path}
\label{sec:path}
The total configuration space to be integrated over is made of elements
$s=[\xx_0,\xx_1,$ $\ldots,\xx_M]$ where $\xx_k$ are the path time slices 
subject to the periodic boundary condition $\xx_M=\xx_0$.
In the simulation we wish to sample these elements from the probability 
distribution,
\bq
\pi(s)=\frac{e^{-{\cal S}}}{Z}~~,
\eq 
where the partition function $Z$ normalizes the function $\pi$ in this 
space.

The idea is to find an efficient way to move the path in a random walk 
sampled by $\pi$, through configuration space.

In order to be able to make the random walk diffuse fast through 
configuration space, as $\tau$ decreases, is necessary to use multislices
moves \cite{ceperley95}.

In our simulation we chose to use the bisection method (a particular 
multilevel Monte Carlo sampling method \cite{ceperley95}). That' s how
an $l$ levels move is constructed.
Clip out of the path $m=2^l$  subsequent time slices $\xx_i,\xx_{i+1},
\ldots,\xx_{i+m}$ (choosing $i$ randomly). In the first level we keep 
$\xx_i$ and $\xx_{i+m}$ fixed and, following L\'{e}vy construction
for a Brownian bridge \cite{levy39}, we move the bisecting point at $i+m/2$
to,
\bq
\xx_{i+m/2}=\frac{\xx_i+\xx_{i+m}}{2}+\mbox{\boldmath $\eta$}
\eq
where {\boldmath $\eta$} is a normally distributed random vector with 
mean zero and standard deviation $\sqrt{\tau m/4}$. As shown in 
next section this kind of transition rule samples the path using a 
transition probability distribution $T\propto \exp(-{\cal S}_f)$. 
Thus we will refer to it as {\it free particle sampling}.
 
Having sampled $\xx_{i+m/2}$, we proceed to the second level bisecting 
the two new intervals $(0,i+m/2)$ and $(i+m/2,i+m)$ generating points 
$\xx_{i+m/4}$ and $\xx_{i+3m/4}$ with the same algorithm. We continue 
recursively, doubling the number of sampled points at each level, 
stopping only when the time difference of the intervals is $\tau$.

In this way we are able to partition the full configuration $s$ into 
$l$ levels, $s=(s_0,s_1,\ldots,s_l)$ where:
$s_0=[\xx_0,\ldots,\xx_i,\xx_{i+m},\ldots,\xx_{M-1}]$, unchanged;
$s_1=[\xx_{i+m/2}]$, changed in level 1;
$s_2=[\xx_{i+m/4},\xx_{i+3m/4}]$, changed in level 2; $\ldots$;
$s_l=[\xx_{i+1},\xx_{i+2},\ldots,\xx_{i+m-1}]$ changed in level $l$. 

To construct the random walk we use the multilevel Metropolis method
\cite{ceperley86,ceperley89,ceperley95}. Call $(s_1^\prime,\ldots,
s_l^\prime)$ the new trial positions in the sense of a Metropolis 
rejection method, the unprimed ones are the corresponding old positions 
with $s_0=s_0^\prime$. 

In order to decide if the sampling of the path should continue beyond level 
$k$, we need to construct the probability distribution $\pi_k$ for level
$k$. This, usually called the {\it level action}, is a function of 
$s_0,s_1\ldots,s_k$ proportional to the reduced distribution function of 
$s_k$ conditional on $s_0,s_1\ldots,s_{k-1}$. The optimal choice for the 
level action would thus be, 
\bq \label{optpi}
\pi^\star_k(s_0,s_1\ldots,s_k)=\int ds_{k+1}\ldots ds_l\,\pi(s)~~.
\eq 
This is only a guideline. Non optimal choices will lead to slower movement 
through configuration space. One needs to require only that feasible paths
(closed ones) have non zero level action, and that the action at the last 
level be exact,
\bq \label{ll}
\pi_l(s_0,s_1,\ldots,s_l)=\pi(s)~~.
\eq  

Given the level action $\pi_k(s)$ the optimal choice for the
transition probability $T_k(s_k)$, for $s_k$ contingent on the levels 
already sampled, is given by,
\bq
T_k^\star(s_k)=\frac{\pi_k(s)}{\pi_{k-1}(s)}~~.
\eq
One can show that $T_k^\star$ will be a normalized probability if and 
only if $\pi_k$ is chosen as in (\ref{optpi}). In general one need to 
require only that $T_k$ be a probability distribution non zero for 
feasible paths. In our simulation we used the free particle transition 
probability of the L\'{e}vy construction as a starting point for a 
more efficient correlated sampling that will be described in a later
section.

Once the partitioning and the sampling rule are chosen, the sampling 
proceeds past level $k$ with probability,
\bq
A_k(s^\prime)=\min\left[1,\frac{T_k(s_k)\pi_k(s^\prime)\pi_{k-1}(s)}
{T_k(s_k^\prime)\pi_k(s)\pi_{k-1}(s^\prime)}\right]~~.
\eq  
That is we compare $A_k$ with a uniformly distributed random number in 
$(0,1)$, and if $A_k$ is larger, we go on to sample the next level. If
$A_k$ is smaller, we make a new partitioning of the initial path, and 
start again from level 1. Here $\pi_0$ needed in the first level can be 
set equal to 1, since it will cancel out of the ratio. 

This acceptance probability has been constructed so that it satisfies a
form of ``detailed balance'' for each level $k$,
\bq \label{db1}
\frac{\pi_k(s)}{\pi_{k-1}(s)}T_k(s_k^\prime)A_k(s^\prime)=
\frac{\pi_k(s^\prime)}{\pi_{k-1}(s^\prime)}T_k(s_k)A_k(s)~~.
\eq
The moves will always be accepted if the transition probabilities and 
level actions are set to their optimal values.

The total transition probability for a trial move making it through all
$l$ levels is,
\bq
P(s\rightarrow s^\prime)=\prod_{k=1}^l T_k(s^\prime)A_k(s^\prime)~~.
\eq
By multiplying Eq. (\ref{db1}) from $k=1$ to $k=l$ and using 
Eq. (\ref{ll}), one can verify that the total move satisfy the 
detailed balance condition,
\bq
\pi(s)P(s\rightarrow s^\prime)=\pi(s^\prime)P(s^\prime\rightarrow s)~~.
\eq
Thus if there are no barriers or conserved quantities that restrict
the walk to a subset of the full configuration space (i.e. assuming 
the random walk to be ergodic) the algorithm will asymptotically 
converge to $\pi$, independent of the particular form chosen for the 
transition probabilities, $T_k$, and the level actions, $\pi_k$ 
\cite{hammersley64}.   
We will call {\it equilibration time} the number of moves needed in the 
simulation to reach convergence.

Whenever the last level is reached, one calculates the properties 
(${\cal K}$ and ${\cal P}$) on the new path $s^\prime$, resets the
initial path to the new path, and start a new move.
We will call Monte Carlo step (MCS) any attempted move.

\section{Choice of $T_k$ and $\pi_k$}
\label{sec:metropolis}
In our simulation we started moving the path with the L\'{e}vy 
construction described in the preceding section. We will now show 
that this means that we are sampling an approximate $T^\star$ with 
free particle sampling.   

For the free particle case $({\cal U}=0)$ one can find analytic expressions 
for the optimal level action $\pi^\star_k$ and the optimal transition 
rule $T_k^\star$. For examples for the first level, Eq. (\ref{optpi})
gives,
\bq
\pi^\star_1(\xx_{i+m/2})&\propto & \rho_f(\xx_i,\xx_{i+m/2};\tau m/2)
\rho_f(\xx_{i+m/2},\xx_{i+m};\tau m/2)\\
&\propto & e^{\frac{1}{m\tau}(\xx_i-\xx_{i+m/2})^2}
e^{\frac{1}{m\tau}(\xx_{i+m/2}-\xx_{i+m})^2}\\
&\propto & e^{\frac{2}{m\tau}\left[\xx_{i+m/2}-\left(\frac{\xx_i+\xx_{i+m}}
{2}\right)\right]^2}~~.
\eq
This justify the L\'{e}vy construction and shows that it exactly samples 
the free particle action (i.e. $A_k=1$ for all $k$'s). This also imply that
for the interacting system we can introduce a {\it level inter action},
$\tpi_k$ such that,
\bq
\tpi_k=\int ds_{k+1}\ldots ds_{l}\,\tpi(s)~~,
\eq
with
\bq
\tpi(s)=\frac{e^{-{\cal U}}}{Z}~~.
\eq
So that the acceptance probability will have the simplified expression,
\bq \label{fs-acceptance}
A_k(s^\prime)=\min\left[1,\frac{\tpi_k(s^\prime)\tpi_{k-1}(s)}
{\tpi_k(s)\tpi_{k-1}(s^\prime)}\right]~~.
\eq

For the $k^{th}$ level inter action we chose the following 
expression,
\bq \label{laction1}
\tpi_k\propto\exp\left[-(\tau\ell_k)^2\sum_{i=1}^{[M/\ell_k]}
\sum_{j=1}^{[M/\ell_k]}V(i\ell_k\tau,j\ell_k\tau)\right]~~,
\eq 
where $\ell_k=m/2^k$. In the last level $\ell_l=1$ and the level
inter action $\tpi_l$ reduces to the exact inter action $\tpi$
thus satisfying Eq. (\ref{ll}).

It' s important to notice that during the simulation we never need to 
calculate the complete level inter action since in the acceptance 
probabilities enter only ratios of level inter actions calculated on the 
old and on the new path. For example if for the move we clipped out 
the interval $t_i,\ldots,t_{i+m}$ with $i+m < M$
\footnote{
When $i+m\ge M$ there is a minor problem with the periodic boundary 
conditions and Eq. (\ref{laction2}) will change.},
we have,
\bq \nonumber
\ln\frac{\tpi_k(s^\prime)}{\tpi_k(s)}=-(\tau\ell_k)^2\left\{\sum_{m=0}
^{2^k}\sum_{n=0}^{2^k}V(t_i+m\ell_k\tau,t_i+n\ell_k\tau)+\right.\\ \label{laction2} 
\left.\sum_{m=1}^{i-1}\sum_{n=0}^{2^k}V(m\ell_k\tau,t_i+n\ell_k\tau)+
\sum_{m=i+m+1}^{M}\sum_{n=0}^{2^k}V(m\ell_k\tau,t_i+n\ell_k\tau)
\right\}~~,
\eq  
which is computationally much cheaper than (\ref{laction1}).

\section{Correlated sampling}
\label{sec:sampling}
\begin{figure}[H] 
\begin{center}
\includegraphics[width=10cm]{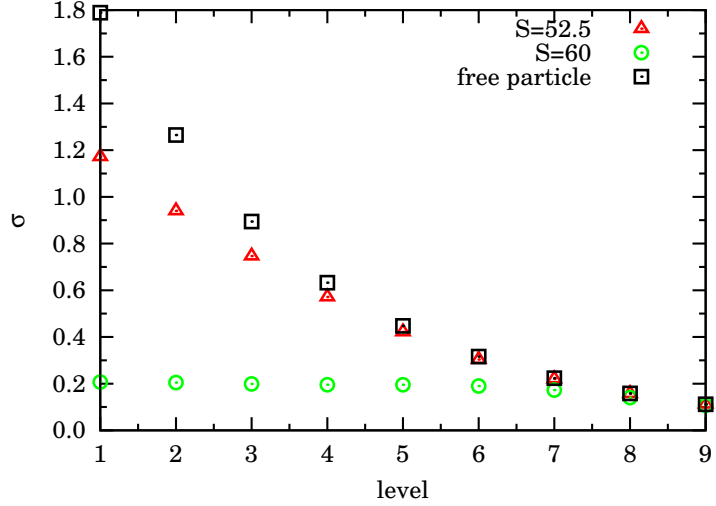}
\end{center}
\caption[deviations]{Shows the deviations (\ref{dev}) for a simulation 
with $S=60$ and $S=52.5$, $\tau=0.025$, $l=9$. The free particle
standard deviations  
(\ref{fp-dev}) are plotted for comparison. For $S=60$ the path is
localized while for $S=52.5$ is unlocalized i.e. closer to the free
particle path.} 
\label{cs}
\end{figure}

When the path reaches equilibrium (i.e. $P(s\rightarrow s^\prime)\approx 
\pi(s^\prime)$) if we calculate,
\bq \label{dev}
\sigma(t_0/\tau)=\sqrt{\left\langle\left[\xx(t)-\left(\frac{\xx(t+t_0)+
\xx(t-t_0)}{2}\right)\right]^2\right\rangle}~~,
\eq 
we see that these deviations are generally smaller than the free particle
standard deviations used in the L\'{e}vy construction (see Fig. 
\ref{cs}),
\bq \label{fp-dev}
\sigma_f(\ell_k)=\sqrt{\ell_k\tau/2}~~.
\eq
As Fig. \ref{cs} shows, the discrepancy gets bigger as $\ell_k$
increases. 

We thus corrected the sampling rule for the correct deviations. For 
example for the first level we used,
\bq
T_1(\xx_{i+m/2})\propto e^{-\frac{\left(\xx_{i+m/2}-\xxb\right)^2}{2\sigma^2
(m/2)}}~~,
\eq
where $\xxb=(\xx_i+\xx_{i+m})/2$. Since the level action is given by,
\bq
\pi_1(\xx_{i+m/2})\propto e^{-\frac{\left(\xx_{i+m/2}-\xxb\right)^2}{2\sigma_f^2
(m/2)}}\tpi_1(\xx_{i+m/2})~~,
\eq 
we can define a function,
\bq
P_1\propto e^{-\frac{\left(\xx_{i+m/2}-\xxb\right)^2}{2}\left[
\frac{1}{\sigma^2(m/2)}-\frac{1}{\sigma_f^2(m/2)}\right]}~~,
\eq
and write the acceptance probability,
\bq
A_1(s^\prime)=\min\left[1,\frac{P_1(s)}{P_1(s^\prime)}\frac{\tpi_1(s^\prime)
\tpi_{0}(s)}{\tpi_1(s)\tpi_{0}(s^\prime)}\right]~~.
\eq
Which is a generalization of Eq. (\ref{fs-acceptance}).

We maintain the acceptance ratios in $[0.15,0.65]$ by decreasing (or
increasing) the number of levels in the multilevel algorithm as the
acceptance ratios becomes too low (or too high).

In the Appendix we report some remarks on the error analysis in our MC
simulations.

\section{Numerical Results}
\label{sec:results}
\begin{figure}[H]
\begin{center}
\includegraphics[width=10cm]{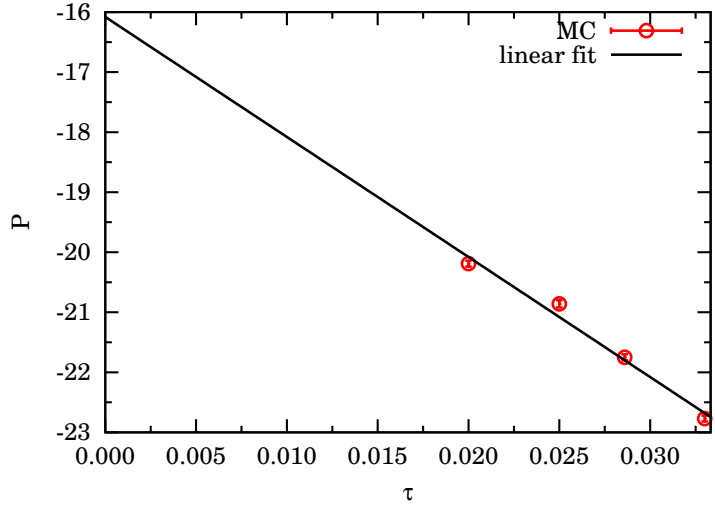}\\
\end{center}  
\caption{Shows the time step, $\tau$, extrapolation for the potential
  energy, $P=\langle{\cal P}\rangle$. We run at $\beta=15,
  \gamma=0.02$, and 
  $S=60$. The extrapolated value to the continuum limit is in this
  case $P=-16.1(5)$ which is in good agreement with the result
  of Ref. \cite{Wang1998}.} 
\label{fig:tau-p}
\end{figure}

\begin{figure}[H] 
\begin{center}
\includegraphics[width=10cm]{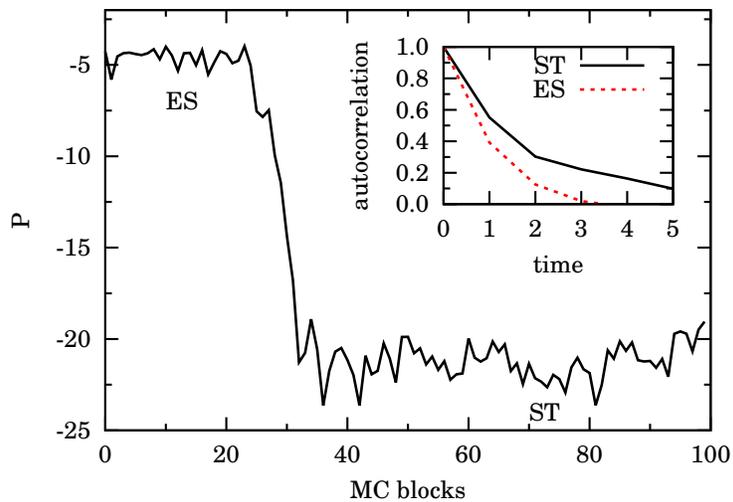}
\end{center} 
\caption{At $S=60$ the results for the potential energy ${\cal P}$ at
  each MC block ($5\times 10^3$ MCS) starting from an initial
  unlocalized path obtained by a previous simulation at $S=52.5$. We
  can see that after about $30$ blocks there is a transition from the ES
  state to the ST state. In the inset is shown the autocorrelation
  function, defined in Eq. (\ref{autocorrelation}), for the potential
  energy, for the two states. The correlation time, in MC blocks, is
  shorter in the  
  unlocalized phase than in the localized one. The computer time
  necessary to carry on a given number of Monte Carlo steps is longer
  for the unlocalized phase.} 
\label{phtr}
\end{figure}

\begin{figure}[H]
\begin{center}
\includegraphics[width=8cm]{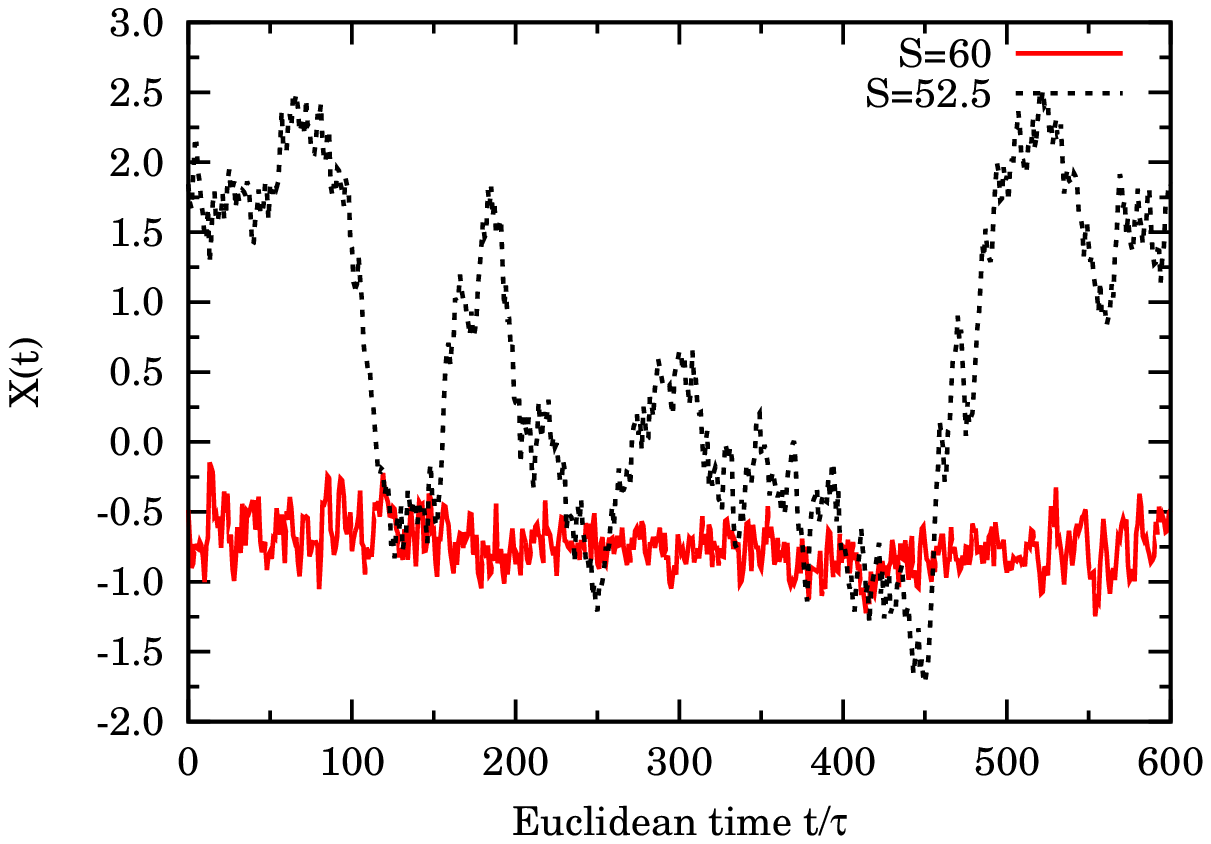}\\
\includegraphics[width=8cm]{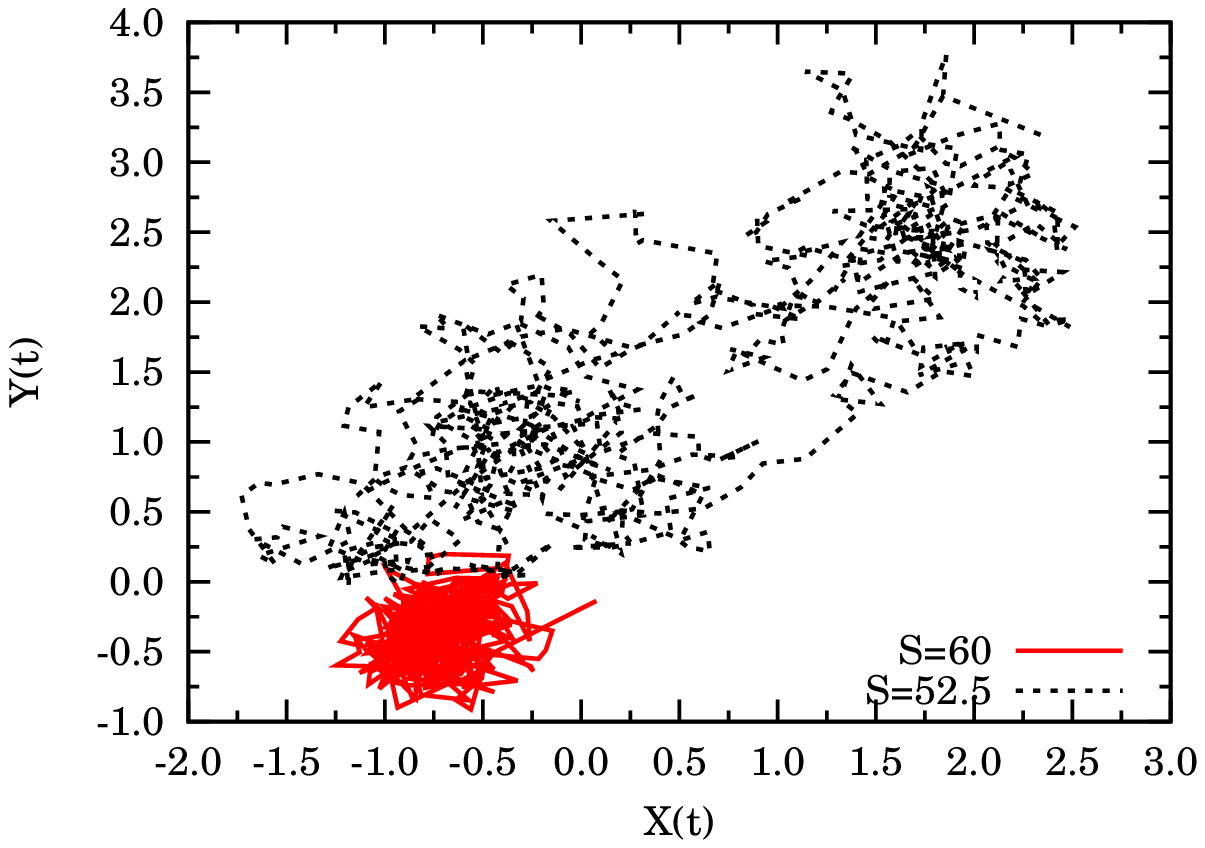}\\
\includegraphics[width=8cm]{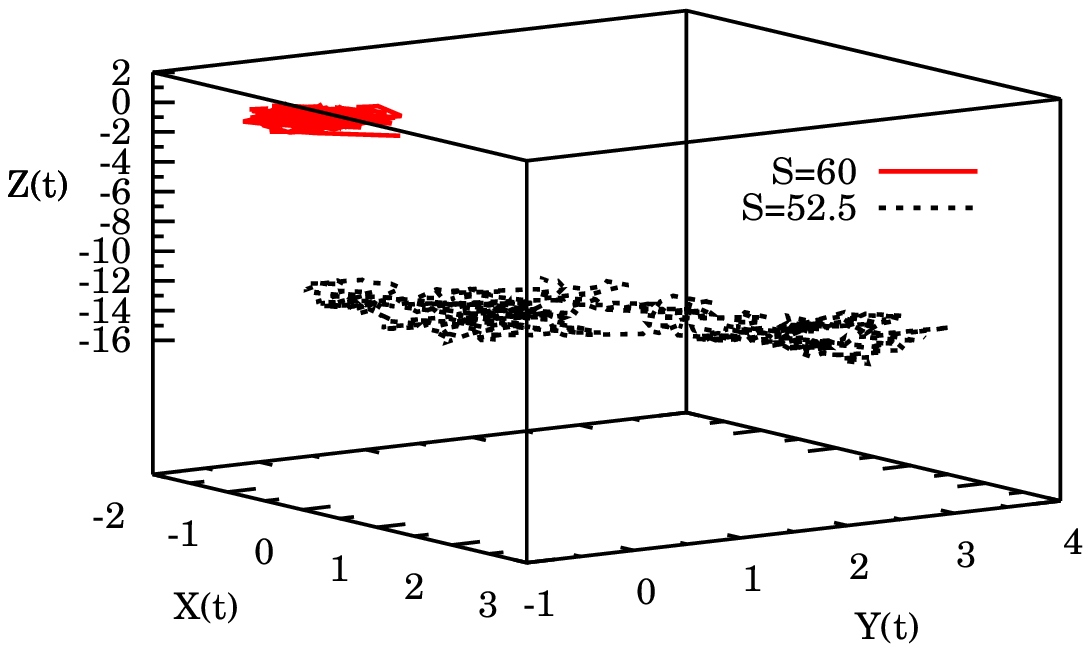}
\end{center}  
\caption{The top panel shows the polaron (closed) path $x(t)$ as a
  function of Euclidean time $t$ in units of $\tau$ at equilibrium
  during the simulation. The
  middle panel shows the projection on the $x-y$ plane of the path. The
  bottom panel shows the three-dimensional path. We see clearly how
  both path has moved from the initial path located on the origin but
  the path at $S=52.5$ is much less localized than the one at $S=60$.}
\label{path}
\end{figure}

\begin{figure}[hbt]
\begin{center}
\includegraphics[width=9cm]{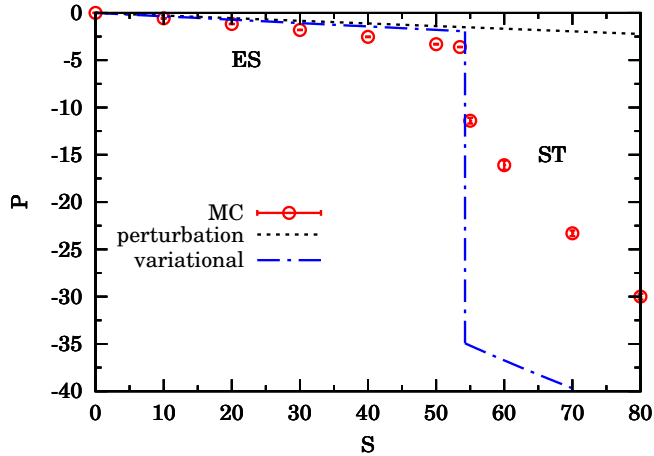}\\
\end{center}  
\caption{Shows the behavior of the potential energy $P$ as a function
  of the coupling constant $S$. The points are the MC results (see 
  Tab. \ref{table}), the dashed line is the second order 
  perturbation theory result (perturbation) valid in the weak coupling
  regime and the dot-dashed line is
  the variational approach from Ref. \cite{sumi73} (variational)
  in the weak and strong coupling regimes.}
\label{fig:s-p}
\end{figure}

We simulated the acoustic polaron fixing the adiabatic
coupling constant $\gamma=0.02$ and the inverse temperature $\beta=15$.
Such temperature is found to be well suited to extract close to ground
state properties of the polaron.
The path was $M$ time slices long and the time step was 
$\tau=\beta/M$. 
For a given coupling constant $S$ we computed the potential energy
$P$ extrapolating (with a linear $\chi$ square
fit) to the continuum 
time limit, $\tau\to 0$, three points corresponding to time-steps
choosen in the interval $\tau\in[1/100,1/30]$. An example of
extrapolation is shown in Fig. \ref{fig:tau-p} for the particular case
$\beta=15, \gamma=0.02$, and 
$S=60$.
In Fig. \ref{fig:s-p} and Tab. \ref{table} we show the results for the
potential energy as a function of the coupling strength. 
With the coupling constant $S=52.5$ we generated the equilibrium 
path which turns out to be unlocalized (see Fig.
\ref{path}). Changing the coupling constant 
to $S=60$ and taking the unlocalized path as the initial path we 
sow the phase transition described in Fig. \ref{phtr}. the 
path after the phase transition is localized (see Fig. \ref{path}). 
Note that since $S$ and $\tau$ appear in the combination $S\tau^2$ in
${\cal U}$ (and $S\tau$ in F) the same phase transition from an ES to
a ST state will be observed increasing the temperature. With the same
Hamiltonian we are able to describe two very different behaviors of
the acoustic polaron as the temperature changes.

In Fig. \ref{fig:s-p} we show the behavior of the potential energy as
a function of the coupling strength. The numerical results suggests
the existence of a phase transition
between two different regimes which corresponds to the so called ES
and ST states for the weak and strong coupling region respectively. We
found that paths related to ES and ST are characteristically
distinguishable. Two typical paths for the ES and ST regimes involved
in Fig. \ref{fig:s-p} is illustrated in Fig. \ref{path}. The path in
ES state changes smoothly in a large time scale, whereas the path in ST
state do so abruptly in a small time scale with a much smaller
amplitude which is an indication that the polaron hardly moves. The local
fluctuations in the results for the potential energy has an
autocorrelation function (defined in
Eq. (\ref{autocorrelation})) which decay much more slowly in the ES
state than in the ST state as shown in the inset of
Fig. \ref{phtr}. Concerning the critical property of the transition
between the ES and ST states our numerical results are in favor of
the presence of a discontinuity in the potential energy.
In the large $\beta$ limit at $\beta=15$ and fixing the adiabatic
coupling constant to $\gamma=0.02$, the ST state appear at a value of
the coupling constant between $S=52.5$ and $S=55$. With the increase
of $\beta$, the values for the potential energy $P=\langle{\cal
  P}\rangle$ increase in the weak coupling regime but descrease in the
strong coupling region. 

From second order perturbation theory (see Ref. \cite{Feynman} section
8.2) follows that the energy shift $E(\gamma,S)$ is given by
$-3S\gamma[1/2-\gamma+\gamma^2\ln(1+1/\gamma)]$ from which one
extracts the potential energy shift by taking
$P(\gamma,S)=\gamma dE(\gamma,S)/d\gamma$. From the
Feynman variational 
approach of Ref. \cite{sumi73} follows that in the weak regime the
energy shift is $-3S\gamma[1/2-\gamma+\gamma\ln(1+1/\gamma)]$ and in
the strong coupling regime $-S+3\sqrt{S/5\gamma}$. 
\begin{table}[hbt]
\caption{MC results for $P$ as a function of $S$ at $\beta=15$ and
  $\gamma=0.02$ displayed in Fig. \ref{fig:s-p}. The runs where made of
  $5\times 10^5$ MCS (with $5\times 10^4$ MCS for the equilibration)
  for the ES states and $5\times 10^6$ MCS (with $5\times 10^5$ MCS
  for the equilibration) for the ST states.} 
\label{table}
{\scriptsize
\begin{center}
\begin{tabular}{||c||c||}
\hline
$S$ & $P$ \\
\hline
\hline
10&  -0.573(8)\\
20&  -1.17(2)\\
30&  -1.804(3)\\
40&  -2.53(3)\\
50&  -3.31(4)\\
53.5&-3.61(1)\\
55&  -11.4(3)\\
60&  -16.1(5)\\
70&  -23.3(3)\\
80&  -30.0(3)\\ 
\hline
\end{tabular}
\end{center}
}
\end{table}
%

\section{Conclusions} 
\label{sec:conclusions}

We used for the first time a specialized path integral Monte Carlo
method to study the low temperature behavior of an acoustic
polaron. At an inverse temperature $\beta=15$ (close to the ground
state of the polaron) and at a non-adiabatic parameter $\gamma =0.02$
typical of ionic crystals we found numerical evidence for a phase
transition between an extended state in the weak coupling regime and a
self-trapped one in the strong coupling regime at a value of the
phonons-electron coupling constant $S=54.3(7)$. The transition also
appears looking at the potential energy as a function of the coupling
constant where a jump discontinuity is observed. Comparison with the
perturbation theory and the variational calculation of
Ref. \cite{sumi73} is also presented.

The specialized path integral Monte Carlo simulation method used as an
unbiased way to study the properties of the acoustic polaron has been
presented in full detail. It is based on the L\'evy construction
and the multilevel Metropolis method with correlated sampling. Some
remarks on the estimation of the errors in the Monte Carlo calculation
are also given in the Appendix. This complement our previous paper
\cite{Fantoni2012} where fewer details on the Monte Carlo method had
been given. 
 
Our method differs from previously adopetd method
\cite{Alexandrou1990,Alexandrou1992b,Crutz1981,Takahashi1983,Wang1998,Kornilovitch1997,Pierleoni2001,Kornilovitch2007}. Unlike
the method of Ref. \cite{Alexandrou1990} our path integral is in real
space rather than in Fourier space,
Refs. \cite{Kornilovitch1997,Kornilovitch2007} put the polaron on a 
lattice and not on the continuum as we did, while
Refs. \cite{Wang1998} use PIMC single slice move whereas the
multilevel PIMC we used instead is a general sampling method 
which can efficiently make multislice moves. The efficiency $\xi$ (see
the Appendix) for the potential energy increases respect to the single
slice sampling because the coarsest movements are sampled and rejected
before the finer movements are even constructed. In
Ref. \cite{Pierleoni2001} the L\'evy construction was used as in our
case but the Metropolis test was performed after the entire path had
been reconstructed, using an effective action, and not at each
intermediate level of the reconstruction. In Ref. \cite{Pierleoni2001}
the simpler Levy reconstruction scheme was also found to be satisfactory
for the efficient sampling of the polaron configuration space even at
strong coupling. Even if we have not implemented the method of
Ref. \cite{Pierleoni2001} we expect our method to be of comparable
efficiency to the one of these authors. Infact it is true that the
Levy construction is computationally cheap but guiding the path as it
is been reconstructed starting already from the first levels as we did
should have the advantage of refining the sampling since the path is
guided through configuration space starting from the small
displacements. 
  
Although our results are of a numerical nature and we only probed the
acoustic polaron for one value of the non-adiabatic parameter $\gamma$
we think that the analysis support the existence of a localization phase
transition as the phonons-electron coupling constant $S$ is increased
at constant temperature or as the temperature is decreased at constant
$S$. More so considering the fact that the introduction of a cut-off
parameter have shown to work successfully in renormalization
treatments. 

\appendix
\section{Estimating errors}
\label{sec:errors}
Since asymptotic convergence is guaranteed, the main issue is whether
configuration space is explored thoroughly in a reasonable amount of 
computer time. Let us define a measure of the convergence rate and of 
the efficiency of a given random walk. This is needed to compare the 
efficiency of different transition rules, to estimate how long the 
runs should be, and to calculate statistical errors.

The rate of convergence is a function of the property being calculated.
Let $\cO(s)$ be a given property, and let its value
at step $k$ of the random walk be $\cO_k$. Let the estimator for the mean
and variance of a random walk with $N$ MCS be,
\bq
O=<\cO_k>=\frac{1}{N}\sum_{k=0}^{N-1}\cO_k~~,\\
\sigma^2(\cO)=<(\cO_k-O)^2>~~.
\eq
Then the estimator for the variance of the mean will be,
\bq
\sigma^2(O)&=&<(\frac{1}{N}\sum_k\cO_k-\frac{1}{N}\sum_k
O)^2>\\
&=& \frac{1}{N^2}<[\sum_k(\cO_k-O)]^2>\\
&=& \frac{1}{N^2}\left\{\sum_k<(\cO_k-O)^2>+2\sum_{i<j}
<(\cO_i-O)(\cO_j-O)>\right\}\\
&=&\frac{\sigma^2(\cO)}{N}\left\{ 1+\frac{2}{N\sigma^2(\cO)}
\sum_{i<j}<(\cO_i-O)(\cO_j-O)>\right\}\\
&=&\frac{\sigma^2(\cO)k_\cO}{N}~~.
\eq
The quantity $k_\cO$ is called the {\it correlation time}
and can be calculated given the autocorrelation function for 
$A_k=\cO_k-O$. The estimator for the {\it autocorrelation
function}, $c_k$, can be constructed observing that in the infinite random walk,
$<A_iA_j>$ has to be a function of $|i-j|$ only. Thus the estimator 
can be written,
\bq \label{autocorrelation}
c_k=\frac{<A_0A_k>}{\sigma^2(\cO)}=\frac{1}{(N-k)\sigma^2(\cO)}
\sum_{n=1}^{N-k}A_nA_{n+k}~~.
\eq
So that the estimator for the correlation time will be,
\bq
k_\cO=1+\frac{2}{N}\sum_{k=1}^N (N-k) c_k~~.
\eq
To determine the true statistical error in a random walk, one needs 
to estimate this correlation time. To do this, is very important that 
the total length of the random walk be much greater than $k_\cO$.
Otherwise the result and the error will be unreliable. Runs in which
the number of steps $N\gg k_\cO$ are called {\it well converged}.

The correlation time gives the average number of steps needed to 
decorrelate the property $\cO$. It will depend crucially on the 
transition rule and has a minimum value of 1 for the optimal rule
(while $\sigma(\cO)$ is independent of the sampling algorithm).

Normally the equilibration time will be at least as long as the 
equilibrium correlation time, but can be longer. Generally the 
equilibration time depends on the choice for the initial path. To 
lower this time is important to choose a physical initial path.
Since the polaron system is isotropic, we chose the initial path 
with all time slices set to $\vec{0}$.

The efficiency of a random walk procedure (for the property $\cO$) is 
defined as how quickly the error bars decrease as a function of the
computer time,
$\xi_\cO=1/\sigma^2(O)N{\cal T}=1/\sigma^2(\cO)k_\cO {\cal T}$
where ${\cal T}$ is the computer time per step. The efficiency depends not 
only on the algorithm but also on the computer and the implementation.



\end{document}